\documentclass[12pt,preprint]{aastex}
\begin{document}
\tighten
\title{Adaptive Optics Echelle Spectroscopy of [Fe II] 1.644 $\mu$m
in the RW Aur Jet: A Narrow Slice Down the Axis of the Flow}

\author{
    Patrick Hartigan \altaffilmark{1},
    and Lynne Hillenbrand\altaffilmark{2}
    }

\vspace{1.0cm}

\altaffiltext{1}{Dept. of Physics and Astronomy, Rice University,
6100 S. Main, Houston, TX 77005-1892, USA}

\altaffiltext{2}{Dept. of Astrophysics, California Institute of Technology,
MC 105-24, Pasadena, CA, 91125}

\begin{abstract}

We present new adaptive optics echelle spectra of the near-infrared [Fe II] lines
in the redshifted and blueshifted jets from the T Tauri star RW~Aur.
The spectra have an unprecedented combination of high
spatial and spectral resolution that makes it possible to trace the dynamics of the flow
to a projected distance of only 10~AU from the source. As noted by previous studies,
the redshifted flow is much slower than its fainter blueshifted counterpart. Our observations
clearly show that both the radial velocities and the emission
line widths are larger closer to the source on both sides
of the jet. The line widths are 20\%\ $-$ 30\%\ of the jet velocity on both sides of the flow,
significantly larger than would be produced by a divergent constant velocity flow.
The observed line widths could arise from a layered velocity structure in
the jet or from magnetic waves.
A bright knot in the redshifted jet has no concomitant increase in line width,
implying that it is not heated by a bow shock.  Alternate heating mechanisms include
planar shocks, ambipolar diffusion and magnetic reconnection.

\end{abstract}


\section{Introduction}

Stellar jets are an integral part of star formation during
the first few million years when an accretion disk is present
around young stars.  How disks generate these
highly collimated supersonic flows is a subject of much current
observational and theoretical work. Most models drive jets
magnetically from the rotating disk or from the point where the
stellar field intersects the inner disk \citep[e.g.][]{fdc06,cai08}, and there is some
evidence that magnetic fields are strong enough within a few hundred
AU of the star to affect the flows \citep{ray97,hartigan07}. The
location where the disk launches material into a jet varies between
models, and observations are just now getting to the
point where they can provide useful constraints on this process
\citep[see][for a review]{ray06}.

The broad goal for observational work related to launching and collimating jets
is to constrain the basic input parameters assumed by the
models. Main outstanding questions involve understanding
how disks accelerate material into jets, how jets become
heated, and what role magnetic fields play in the flow
dynamics. To answer these questions requires observations as close as
possible to the star, because as the flow evolves velocity variability in
the jet creates internal bow shocks that dominate the heating and 
complicate the internal dynamics.
If we wish to infer the shapes of streamlines and the location and strength of shock waves,
and detect any rotational component to the outflow we must measure precise
radial velocities and resolve emission line widths down to the thermal broadening limit.

The observations described above are near or beyond the capabilities of
existing facilities. The closest bright stellar jets are $\sim$ 140~pc away,
so to study these systems on scales of 10~AU requires spatial resolutions
$\lesssim$ 0.1$^{\prime\prime}$. The strongest
emission lines from jets are those in the optical ([S~II], [N~II], [O~I], H$\alpha$, etc).
However, in the optical, the required spatial resolution is available only
from HST, and unfortunately current HST instrumentation delivers only moderate
spectral resolution which fails to reach the thermal broadening limit by
a factor of five or more.  The only atomic near-infrared lines of note in stellar jets are 
Fe [II] 1.644 $\mu$m and 1.667 $\mu$m. Ground-based adaptive optics systems on
large telescopes reach the required spatial resolution and possess
instruments that also deliver the necessary spectral resolution to study these lines
in a few of the brightest jets.  It is also becoming possible to
study molecular jets close to their sources using radio interferometric
observations of SiO \citep[e.g.][]{cabrit07,codella07}.

Because it is bright, relatively close ($\sim$ 140 pc), and visible all
the way to the star, the RW Aur jet is one of the best targets in the
sky to study how accretion disks launch jets.
RW~Aur was discovered as a variable star a little over a century
ago \citep{oldref}, and is well-known as having a rich emission line
spectrum in the optical characteristic of a high-accretion rate classical
T~Tauri star \citep{joy45,hartigan91,sbbb00}.  The system is also a binary, with
a companion located $\sim$ 1.4$^{\prime\prime}$ from the primary
\citep{joy45,wg01}.  The jet has both redshifted and blueshifted
components, with the redshifted part having higher radial velocity
\citep{hirth94,hamaan94}.  The jet and its bow shocks extend
from the star for at least 100$^{\prime\prime}$ \citep{mcgroarty04},
and proper motions show the jet to be inclined 46 degrees to the line of sight
\citep{lopez03}.

Several studies of RW Aur have focused on the properties of its jet
on subarcsecond scales as it emerges from the source.
High-resolution optical images of [O~I] and [S~II] 
traced the jet to within $\sim$ 0.3 arcseconds, of the star
\citep{dougados00}, and showed that it has a roughly constant opening
angle of 3.9$^\circ$. The jet does not project back to a point at the star,
indicating that the opening angle is wider within 30~AU of the source.
Spectra of the jet from STIS on HST
show that the density declines with distance from the star as the
jet expands, and that the velocity asymmetry between the
redshifted and blueshifted jets has increased over the last decade
\citep{woitas02}. Emission-line spectroastrometry provides evidence for
outflows on AU scales \citep{whelan04}, and the variability of
high-velocity components of semiforbidden UV lines also implies 
a wind close to the star \citep{gdc03}, consistent with recent models of
the Balmer line profiles \citep{alencar05}.

An intriguing observation of asymmetry in the velocity structure of
the jet has been interpreted as evidence for rotation \citep{coffey04,woitas05}.
However, these observations are controversial because the modest spectral
resolution of STIS (55 km$\,$s$^{-1}$) makes it very difficult
to infer rotational velocities of only a few km$\,$s$^{-1}$, and in the case
of RW~Aur the molecular disk rotates in the
opposite sense of the inferred jet rotation \citep{cabrit05}.
Precession can also mimic rotation signatures in outflows
\citep{cerqueira06}.

Two high-spectral resolution studies of the infrared [Fe~II] lines in RW~Aur
have been done.  \citet{davis03} used an echelle with 19 km$\,$s$^{-1}$
resolution and a wide slit without adaptive optics to resolve the line
widths in the redshifted and blueshifted jets. More recently,
\citet{pyo06} used adaptive optics with the Subaru telescope 
with 0.2$^{\prime\prime}$ resolution, a 0.3$^{\prime\prime}$ wide slit and
30 km$\,$s$^{-1}$ velocity resolution to study the kinematics of the jet
down to $\sim$ 30~AU from the source. Such observations provide 
important constraints on models of jet collimation and energetics.

In this paper, we present the first high resolution adaptive optics observations
of a stellar jet taken with the near-IR spectrometer on Keck.
These observations represent a significant improvement in the
spatial and spectral resolution over previous studies - 
a factor of $\sim$ 3 better spectral resolution and
20\% better spatial resolution than the HST STIS data, and
a factor of 1.5 and 3 better spectral and spatial resolution, respectively,
than the Subaru spectra. In addition, the slit width of 0.068$^{\prime\prime}$
is $\sim$ 4 times narrower than that of the Subaru spectra, so we can isolate
a slim cross section of the jet. 
Both the spatial and spectral dimensions are subsampled a factor of $\sim$ 4
better than those of STIS, which enables more precise subtraction of the stellar
PSF close to the star.

The cost to pay for such high spatial and spectral resolution and a very
narrow slit is a reduction in flux.
We were unable to detect the fainter extended emission away from the
axis of the RW Aur jet, and thus cannot address issues related to rotation.
However, our observations clearly resolve the line widths and velocities
in the jet for the first time to within 10 AU of the star. The narrow
slit provides a sharp `cut' down the axis of the flow which is easy
to interpret, and makes it possible to measure the geometry of the streamlines as
the jet emerges from the disk.

\section{Observations}

We obtained spectra of the RW Aur jet with
the NIRSPEC spectrograph and adaptive optics 
system attached to the W.~M.~Keck~II telescope on Mauna Kea on 17 Dec 2006 UT.
The position angle of the slit was 125 degrees (along the axis of the jet), and the
AO system used the target, RW Aur, as a natural guide star (R magnitude $\sim$ 10).
The slit width was 68 mas, which projects to a distance of only
9.5~AU at the distance to the object of 140~pc, so these observations
represent narrow slices down the axis of the flow.

Spectra include seven orders from 1.51 $\mu$m through 1.74 $\mu$m, with 
gaps of $\sim$ 0.03 $\mu$m where the spectra extend
beyond the 1024$\times$1024 InSb chip.
The dispersion was 0.238 \AA\ (4.34 km$\,$s$^{-1}$) per
pixel at the wavelength of the [Fe II] 1.644 $\mu$m emission line.
The velocity resolution of the instrument, defined by the FWHM of
several OH night sky emission lines, was 20.3 km$\,$s$^{-1}$ and
remained constant throughout the night. The dispersion is about
five times finer and the resolution about three times better than
optical spectra of RW~Aur obtained with STIS on HST \citep{woitas02, woitas05}.

Wavelength shifts were less than 0.1 pixel between different sky spectra,
so a single wavelength solution suffices for all spectra. We determined the
wavelength scale from positions of OH emission lines in our night sky spectra
\citep{rousselot99}, and from Kr and Ar lines in calibration lamp spectra.
The number of lines per order is sparse, so uncertainties in the
wavelength solutions are rather large, $\sim$ $\pm$ 5 km$\,$s$^{-1}$ for 1-sigma.
The wavelength calibration is most uncertain at the edges of orders. This
uncertainty affects the velocity zero point of our spectra but has
a negligible effect on the line profile shape or the velocity dispersion.

Flatfields taken with the continuum lamp showed an irregular interference
pattern that introduced spurious patterns to our data. The chip is quite flat,
with variations in pixel-to-pixel sensitivity only 1\%\ $-$ 2\% , so
in lieu of a continuum flat we used the dithers of the multiple exposures along
the slit to remove pixel-to-pixel variations. The PSF is well-sampled
both spatially and spectrally, further reducing the need for continuum flats.

Individual exposure times were five minutes in length. We initially
acquired spectra in a grid centered on the RW~Aur jet with
offsets perpendicular to the axis of the jet at distances
that ranged from $-$0.204$^{\prime\prime}$ to $+$0.204$^{\prime\prime}$,
spaced in intervals of one slit width, 0.068$^{\prime\prime}$.
However, observations with offsets from the axis $\ge$ 0.1 arcsecond
showed no signal from [Fe~II] so we resampled the interior part of
the grid with 0.034$^{\prime\prime}$ spacings, and concentrated most of our 
spectra within $\pm$ 0.1 arcsecond of the axis.
In all, 27 exposures had some [Fe II] signal.
Nine additional spectra of blank sky taken throughout the night
aided the wavelength calibration and background subtraction of
OH night sky lines. 

In practice, small spatial offsets exist between the desired and
actual slit positions. Fortunately, these are easy to
quantify with the 256x256 HgCdTe slit camera (SCAM) images, which are
taken simultaneously with the spectra.  RW Aur and RW Aur/c 
appear on all SCAM images, and define the SCAM plate scale to be 16.3 mas
per pixel, close to the value quoted by
\citet{hornstein02}. The slit is visible on the SCAM images, and
is 2.20$^{\prime\prime}$ in length. On the science camera,
spectral images of the slit in 
the night sky emission lines subtend 118 pixels,
so the plate scale of the longslit spectra is 18.6 mas per pixel. 

Conditions were non-photometric, but SCAM images of RW Aur/c show
the AO system produced stellar images with FWHM of 60 $\pm$
13 mas at H throughout the night (Table~1). This spatial resolution is
about 20\%\ better than that obtained with optical diffraction-limited dithered
HST slitless images of the HH~30 jet with STIS \citep{hep04}, and a factor of
three better than the Subaru spectra of \citet{pyo06}.
While some previous observations
have suggested that the companion may also be a binary, our observations
show only a single star, in agreement with \citet{wg01}.

We combined the [Fe~II] spectra
into five position-velocity diagrams as described in Table~1 and Fig.~1, 
using the position of RW Aur/c in the SCAM
images to define where the slit was positioned during
each exposure. The SCAM images also show that there were 
shifts of up to 0.10 arcsecond along the slit, which we removed before
averaging individual frames into the five output spectra,
denoted 0.083-NE, 0.033-NE, Center, 0.047-SW, and 0.089-SW.
After subtracting a background sky image from each object and removing 
any remaining hot pixels, we extracted the order
of interest and corrected the spectra for tilts in the x- and y-directions
using the observed trace of the RW Aur spectrum and the position of the
night sky lines, respectively.  The reduced data have spatial position
along the y-axis and heliocentric radial velocity along the x-axis.
The heliocentric radial velocity of RW~Aur is 14 $\pm$ 5
km$\,$s$^{-1}$ \citep{hartmann86}.

\section{Analysis and Results}

The method we use to extract extended emission lines from longslit spectra
is that of \cite{hep04}. Briefly, the spatial PSF of the star
is defined by coadding the counts in a wavelength region on either 
side of the emission line, and the PSF is interpolated linearly between these
regions. Once one obtains a spectrum at the location of the
star, multiplying this spectrum by the spatial PSF and subtracting it from the
image leaves only the extended line emission. To obtain a stellar spectrum
we split the spectrum into redshifted and blueshifted portions. The blueshifted
flow occurs to the SE, so for the blueshifted portion of the spectrum
we extracted the stellar spectrum from the NW part of the stellar PSF.
Similarly, the SE part of the PSF defined the spectrum of the redshifted portion
of the spectrum. Combining the two halves gives a position-velocity diagram
for the extended line emission. The procedure is implemented through an
imfort program in IRAF\footnote{IRAF is distributed by the
National Optical Astronomy Observatory, which is operated by the
Association of Universities for Research in Astronomy (AURA)
under cooperative agreement with the National Science Foundation.}.

Figure~1 shows that the position-velocity diagrams of the five positions
are essentially identical. We could not detect any flux at
distances greater than 0.1 arcsecond away from the axis of the flow. Combining
the AO-corrected FWHM of 60 mas in quadrature with the slit width of 68 mas we
obtain $\sim$ 90 mas for the expected spatial extent of an unresolved source.
Hence, our observations do not clearly resolve the spatial width of the jet.
HST-STIS spectra of [S~II] with comparable spatial resolution to our spectra
do resolve the jet spatially \citep{woitas02}, but with a much lower spectral
resolution. 

We measured the radial velocities and FWHM from each 5-minute exposure
individually and combined the results to produce each point and errorbar in
Fig.~2 and Fig.~3.  The velocity of the redshifted flow is almost a factor of
two lower than that of the blueshifted flow, as observed already by
\citet{pyo06}, and by \citet{hirth94}. The combination of
exquisite spatial and spectral resolution of the Keck AO system 
reveals that the radial velocity of the redshifted 
[Fe II] decreases in the first 0.2$^{\prime\prime}$, and then stays constant.  
The redshifted emission is slower, brighter, and more peaked in velocity
than the blueshifted emission.

Fig.~3 shows that the [Fe~II] line is well-resolved spectrally,
and its width increases monotonically in both the jet and the
counterjet as the distance to the source decreases.
The blueshifted flow has
a linewidth about 50\%\ larger than its redshifted counterpart.
The larger linewidth of the blueshifted flow was noted before by
\citet{pyo06}, though that work did not have sufficient spatial 
and spectral resolution to detect the rise of linewidth within
$\sim$ 30~AU of the source within the redshifted jet.

RW~Aur has many permitted emission lines in its spectrum \citep{petrov01},
and several of the Brackett lines appear in our observations.
A good way to determine whether or not these lines have a wind component is
to perform spectroastrometry, which determines spatial offsets of the emission
lines with respect to the adjacent continuum \citep[e.g.][]{ray06}.
While the method is powerful because it is very sensitive to small positional
shifts and can be done as a function of velocity for resolved emission lines,
it suffers from the important limitation that the weighted average of
the line emission at a given velocity determines the amount of the spatial
shift. Hence, if, for example, a blueshifted part of the line profile has
a centroid shifted 2~AU from the star, there need not be any
emission at 2~AU. Such a shift only implies that the weighted average of the
emission from the star and in the flow is offset from the stellar position
by 2~AU at that radial velocity. 

Fig.~4 presents spectroastrometry in the vicinity of the bright Brackett
and [Fe~II] lines. The spectra show no indications of any wind signatures
in Br-12 or Br-11 at the 0.04 pixel level ($\sim$ 0.1~AU). However, both
the [Fe~II] 1.644 $\mu$m line and the much weaker [Fe~II] 1.677 $\mu$m
line exhibit clear redshifted and blueshifted astrometric signatures
even though the line emission is dominated by photospheric and veiling flux
at this wavelength.

\section{Discussion}

\citet{fdc06} summarize three proposed geometries for magnetohydrodynamic jets in which
open field lines are anchored to rotating young stellar objects.
The models are all variations of the magnetocentrifugal mechanism \citep{bp82},
but differ in their mass loading: the disk wind model \citet{kp00} typically loads material
over a wide range of radii; the X-wind model \citep{shang02} drives the wind from a
narrow range of radii; and the hybrid stellar wind model \citep{mp05}
emanates from latitudes between the base of the accretion column and the rotational pole. 
In each case the winds are accretion-powered and the models can be
``cold" or ``warm" depending on whether there is added thermal energy. 

Ideally we would like to use the position-velocity diagrams in Fig.~1 to
test models of outflows directly.  However, to do so requires that models 
accurately predict radiation from the [Fe~II] 1.644 $\mu$m line, which in turn requires
that the models incorporate all the important heating and cooling processes in
their calculations and can predict the flux in this emission line.
This is a tall order for any model, because time-variable phenomena such as internal
velocity variability and magnetic reconnection appear to dominate the heating in jets
within a few hundred AU of the source.  For example,
recent HST observations of the physical conditions in the
HH~30 jet show that it emerges mostly neutral,
becomes ionized several tens of AU from the source, and bright knots
then move downstream in the flow \citep{hm07}.
In RW~Aur there is a large velocity difference between the redshifted and blueshifted jets,
with the redshifted jet the brighter of the two.  Steady-state, axially-symmetric
models cannot account for any of these phenomena, which ultimately determine how
bright lines appear in position-velocity diagrams.  Reconnection geometries in stellar
jets depend upon whether the magnetic axis of the disk aligns with that of the star \citep{fdc06},
but detailed calculations of the position-velocities produced by this type of heating
are lacking.

Hence, the most robust comparison we can make at present with model position-velocity diagrams
is with the velocity extent of the line profile at various places along the jet.  This comparison
is not without flaws, because if the line does not emit in a particular part of the flow, as might
occur if the gas was too cold there or had a different ionization state than the line of interest,
then the line profile will not sample the kinematics of that region. In the case of the RW~Aur jet,
the gas is unlikely to be Fe~I, which is easily ionized, or more ionized than Fe~II because no
other high ionization lines like [O~III] exist close to the star. However, the near-IR [Fe~II] lines
begin to become collisionally quenched when electron densities like those in the jet close to the
source ($\gtrsim\ 10^4\,$cm$^{-3}$) are reached, and the emission will also decline
if the gas cools below a few thousand Kelvin.

Both the redshifted and blueshifted jets have similar
appearances in the position-velocity diagram, in that the line profile is wider
near the source and narrows with distance. This general behaviour occurs in all MHD launching
scenarios as the jet becomes collimated, though, as expected, none of the models reproduce the bright knot
located about 30~AU from the source on both sides of the flow.  One potential interesting constraint 
from Fig.~1 is that no blueshifted emission occurs on the redshifted side of the jet, and no redshifted
emission exists on the blueshifted side. This is in conflict with at least one
X-wind model \citep{shang98,pyo06}, where a wide opening angle near the source produces 
a large radial velocity dispersion. However, this objection disappears if the jet is simply
heated at a larger distance once the flow becomes more collimated. 
The X-wind model has a relatively narrow linewidth like that observed, but a
disk wind can also produce similar linewidths, depending on the model parameters
\citep[e.g.][]{pesenti04}.

As noted by previous authors \citep[e.g.][]{woitas02} the fact that the redshifted
jet is visible close to the star implies that the circumstellar disk does not
block that portion of the flow. Our spectra trace the redshifted jet to within $\sim$ 10~AU
from the star, so using the known inclination i = 46$^\circ$ for RW~Aur \citep{lopez03}, any opaque
disk cannot extend beyond $\sim$ 15~AU from the star. This constraint does not
seem particularly surprising, especially since RW~Aur is a known binary and the
companion could play a role in truncating the outer disk of the primary through dynamical
interactions \citep[e.g.][]{bate09}.

As noted above, both the redshifted and blueshifted flows have
what appears to be a bright knot at about 30~AU from the source
(Fig.~1).  Without time-resolved observations we cannot determine if these features are
stationary, or, as typically seen in other flows \citep[e.g. HH~30][]{hm07},
move with the jet. Some of the decline of [Fe~II] flux close to 
the source on the redshifted side may be caused by the disk blocking our view.
The knots are not accompanied by a sudden increase in linewidth, as one would expect for
a strong bow shock, but these features could still represent shock waves if the
shocks were planar because the curved nature of bow shocks is what produces 
large line widths. It is worth noting that this close to the source the
magnetic field strength may be large enough to significantly inhibit shock formation
from variable flow velocities \citep{hartigan07}, so the jet may be heated by
other mechanisms such as magnetic waves or ambipolar diffusion at tens of AU
from the star.

We can use the observed deconvolved line widths to infer something about the flow
geometry, assuming radial flow (rotation signatures are at most a few 
km$\,$s$^{-1}$, and thermal broadening for Fe is also of this order). 
The ratio of the linewidth to the jet velocity is
an easy dimensionless quantity to measure that depends on the flow geometry. The
velocity along the axis of the jet is V$_{jet}$ = V$_{rad}$ / cos(i), where V$_{rad}$
is the observed radial velocity.
The linewidth along a slice through the axis of the jet
caused by streamlines that diverge at a full opening angle of $\delta$i is simply
$\Delta$V = V$_{jet}$ sin(i) $\delta$i so that $\Delta$V / V$_{jet}$ = sin(i) $\delta$i.
For RW~Aur, i = 46 $^\circ$, sin(i) $\sim$ 0.7 and $\delta$i $\sim$ 4$^\circ$ so that $\Delta$V / V$_{jet}$
$\sim$ 0.05.  Fig.~5 shows that the observed value of $\Delta$V / V$_{jet}$ (corrected for
instrumental resolution) for both the redshifted and blueshifted portion of
the flow is much larger, $\sim$ 0.25.  Hence, the large linewidth in the jet is dominated by
different velocities within the slit and not by projection effects of an expanding
radial flow with a single velocity. As noted above, the lack of correlation between intensity
and line width suggests bow shocks play a minor role in altering the line width.

In the above calculation we have assumed 4$^\circ$ as the opening angle, as observed by
\citet{dougados00} at distances down to 30~AU from the source, about the location of 
peak redshifted and blueshifted emission in Fig.~1.
At closer distances a larger opening angle would increase the linewidth for a radial flow. 
However, Fig.~1 shows that the linewidth increases only about 10\% between 30~AU and $\sim$
10~AU, where the emission becomes lost in the stellar profile, so the effect of a larger
opening angle on the linewidth between 30~AU and 10~AU appears to be negligible. This is
not surprising, as there is no evidence for a larger opening
angle in this distance range for other YSO jets that have been observed with STIS
\citep{hep04}.

Because the observations were taken with such a narrow slit, and no emission was
observed away from the axis of the flow, the line emission
we detect originates from $\sim$ 14~AU of the axis.
According to Fig.~5, within this volume the velocity of the [Fe~II] emitting gas must change by $\sim$ 25\%.
In a disk wind model the velocity would need to drop off quickly enough off-axis to produce this
spread in line width.  Alternatively, if the jet is heated in
some way by magnetic waves, then the line width should be
on the order of the Alfven speed in the jet.
A density of $10^5$ cm$^{-3}$ and a linewidth of 70 km$\,$s$^{-1}$ implies
a field of $\sim$ 10~mG if the broadening is purely magnetic.
The Alfvenic Mach number of the flow tens of AU from the star
would be V$_{jet}$ / $\Delta$V $\sim$ 4.

Fig.~5 shows two other interesting aspects of the flow. First, despite the large radial velocity
differences between the redshifted and blueshifted jets, the ratio $\Delta$V / V$_{jet}$
is remarkably similar between the two sides of the flow. Second, there is a subtle, but significant
trend that this ratio decreases with distance from the star. This decrease occurs because 
the linewidth drops more rapidly with distance (Fig.~3) than the radial velocities do (Fig.~2). 
In the context of pure magnetic heating this trend would imply a gradual increase of the
magnetosonic Mach number with distance, as one would expect if reconnection or ambipolar diffusion
were responsible for heating the jet.

\section{Summary}

Our Keck spectra of the RW~Aur jet are a significant improvement in both spatial
and spectral resolution over previous studies of any stellar jet. The jet was unresolved
perpendicular to the flow, but the new spectra provide valuable information about the kinematics
of the jet to within 10~AU of the source. We confirm several features reported previously,
including the velocity asymmetry and bright redshifted flow close to the source. Our observations
clearly show larger line widths close to the source for both the blueshifted and redshifted
sides of the flow.  The ratio of the line width to the jet velocity
is about 0.25 for both sides of the flow, a factor of 5 greater than should occur from projection effects
in a radial flow. The observed line widths could arise from a layered velocity structure in
the jet or from magnetic waves.
Spectroastrometry shows no evidence for outflow in the upper Brackett lines.

The Keck spectra illustrate what should become possible as instrumentation on large telescopes
continues to improve. Ideally one would like to have the same or better spectral and spatial resolution
as these observations but with a system sensitive enough to detect emission away from the axis
of the flow in order to address issues such as rotation and to better quantify the 
dynamics of the flow. Time-resolved observations are also crucial for determining if
the observed knots are stationary or move outward with the jet. If AO systems could be made
to work in the optical where emission lines are bright and diagnostics are plentiful we
could also learn a great deal about how jets are heated as they emerge from their disks.

\acknowledgements{PH was supported by NASA Origins grant NNG05GH97G during the course of this work.
We thank Suzan Edwards for her comments on a draft of this paper.
}

\null
\begin{center}
\begin{deluxetable}{cccc}
\singlespace
\tablenum{1}
\tablewidth{0pt}
\tablecolumns{4}
\tabcolsep = 0.04in
\parindent=0em
\tablecaption{Positional Offsets of RW Aur Slit Spectra}
\startdata
\noalign{\medskip}
\noalign{\medskip}
\noalign{\hrule}
\noalign{\medskip}
Sequence Number&Spatial Offset$^a$ &FWHM of PSF$^b$&Output Image$^c$ \\
\noalign{\smallskip}
\noalign{\hrule}
\noalign{\smallskip}
64 & -0.099  &0.065 & 0.083 NE\\
39 & -0.093  &0.064 &         \\
59 & -0.078  &0.042 &         \\
26 & -0.062  &0.083 &         \\
\\
65 & -0.044  &0.052 & 0.033 NE\\
33 & -0.039  &0.080 &         \\
75 & -0.024  &0.049 &         \\
76 & -0.024  &0.049 &         \\
77 & -0.024  &0.049 &         \\
\\
50 & -0.008  &0.049 &  Center \\
51 & -0.008  &0.049 &         \\
52 & -0.008  &0.049 &         \\
53 & -0.008  &0.049 &         \\
82 & -0.008  &0.046 &         \\
55 &  0.000  &0.072 &         \\
56 &  0.000  &0.072 &         \\
57 &  0.000  &0.072 &         \\
58 &  0.000  &0.072 &         \\
30 &  0.003  &0.088 &         \\
66 &  0.029  &0.054 &         \\
\\
78 &  0.044  &0.046 &  0.047 SW\\
79 &  0.044  &0.046 &          \\
80 &  0.044  &0.046 &          \\
35 &  0.057  &0.070 &          \\
\\
22 &  0.085  &0.073 &  0.089 SW\\
72 &  0.086  &0.051 &          \\
67 &  0.096  &0.070 &          \\
\\
\enddata
\tablenotetext{a}{Spatial offset perpendicular to the axis of the jet.
Positive values denote offsets to the SW, and negative values are offsets
to the NE. The position angle of the slit on the sky was 125 degrees.}
\tablenotetext{b}{FWHM in arcseconds of the SCAM image of RW Aur/c.}
\tablenotetext{c}{Each spectrum was averaged into one of five output images.}
\end{deluxetable}
\end{center}
\vfill\eject

\begin{figure}

\vbox to 5.0in{\includegraphics{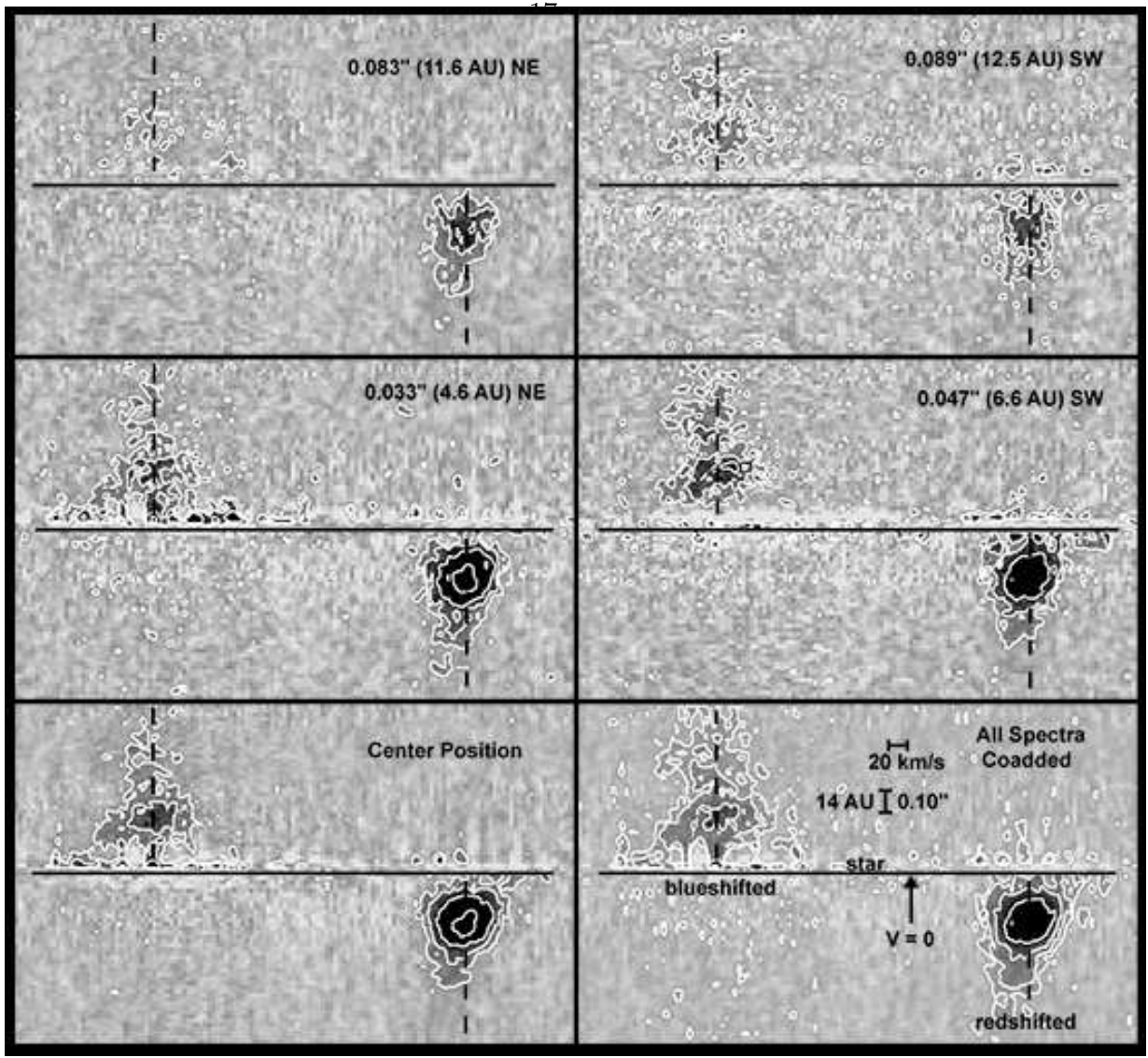}}
\caption{Position-velocity diagrams of the [Fe~II] 1.644 $\mu$m emission line
for five slit positions along the RW Aur jet (PA = 125 degrees),
coadded as described in the text. The velocity resolution is 20 km$\,$s$^{-1}$
and the spatial resolution $\sim$ 0.06 arcseconds. Thin horizontal black lines
in each frame show the spatial location of the star, and the arrow in the bottom
right frame points to zero (heliocentric) velocity.
Dashed vertical lines mark the velocities of
the blueshifted and redshifted flows. Adjacent contours differ by a factor of two.
There are no significant differences
in the position-velocity diagrams for the different slit positions, which are
sampled at roughly the Nyquist frequency of the instrumental resolution.
The redshifted jet is brighter and slower than the blueshifted jet.
}
\end{figure}
\vfill\eject

\begin{figure}
\vbox to 5.0in{\includegraphics{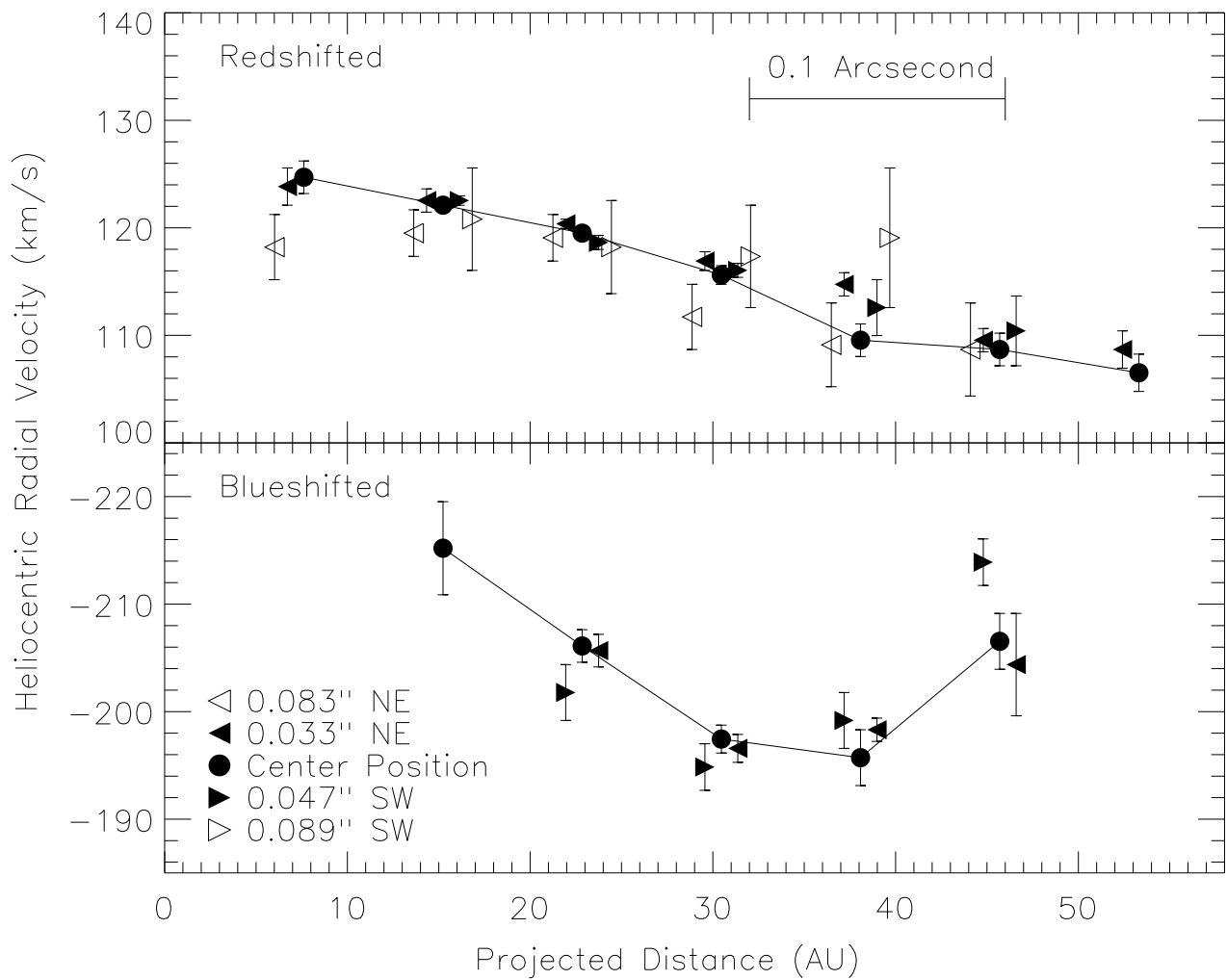}}
\caption{Radial velocities of the [Fe II] 1.644 $\mu$m emission for the redshifted
and blueshifted parts of the flow. Points taken at the same projected distance
for the different slits
are offset from one another slightly to avoid confusion. 
Scatter of values obtained from each 5-minute exposure produces the 1-sigma errorbars. 
Both the redshifted
and blueshifted parts of the flow slow down as the jet emerges from the source.
There are no significant differences in the radial velocities observed
in the different slit positions.
}
\end{figure}
\vfill\eject

\begin{figure}
\vbox to 5in{\includegraphics{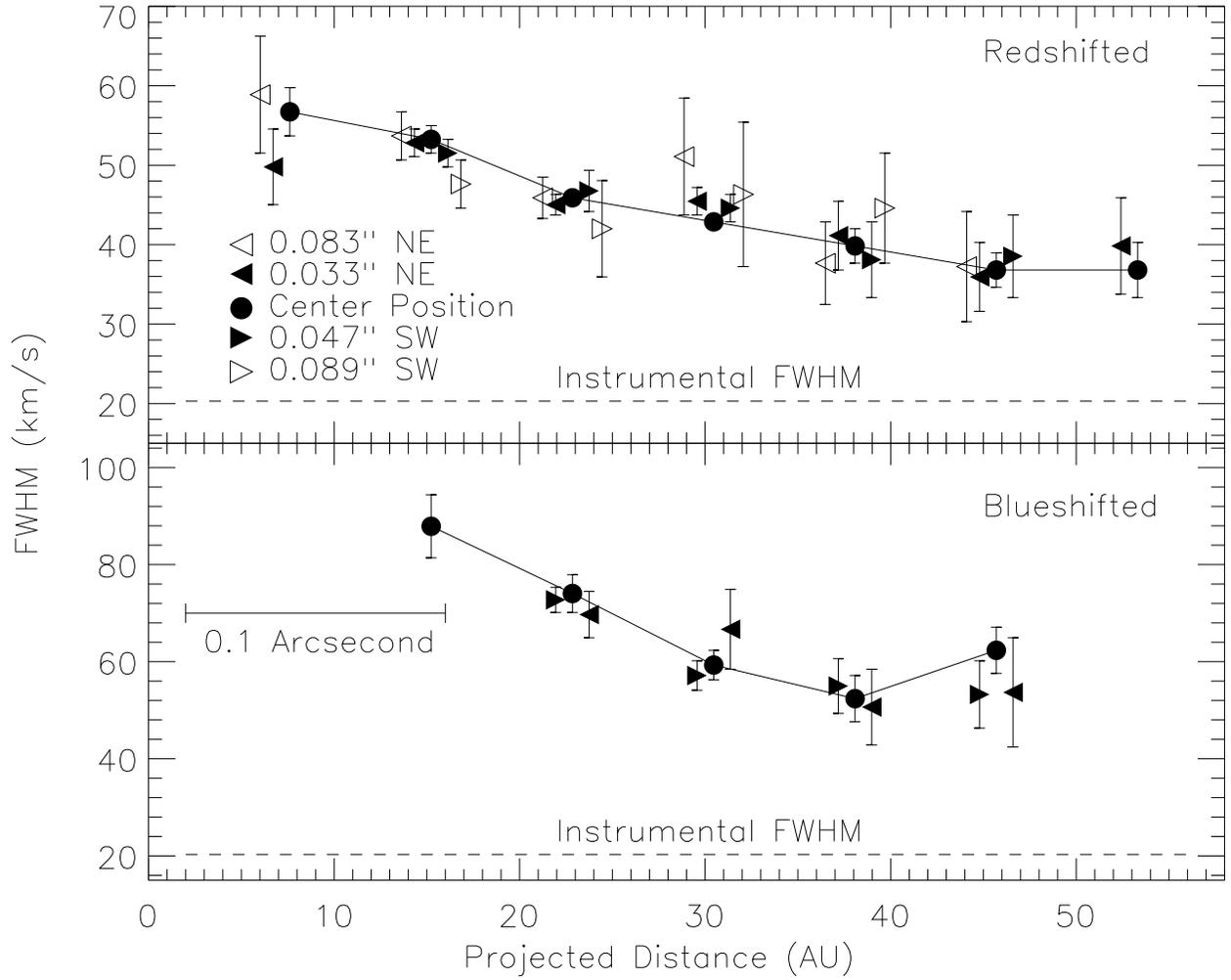}}
\caption{Same as Fig.~2 but for the FWHM of the [Fe II] 1.644 $\mu$m
line profiles. The emission line widths are resolved for all points in the figure, and
decrease monotonically with distance from the star.
}
\end{figure}
\vfill\eject

\begin{figure}
\vbox to 5.0in{\includegraphics{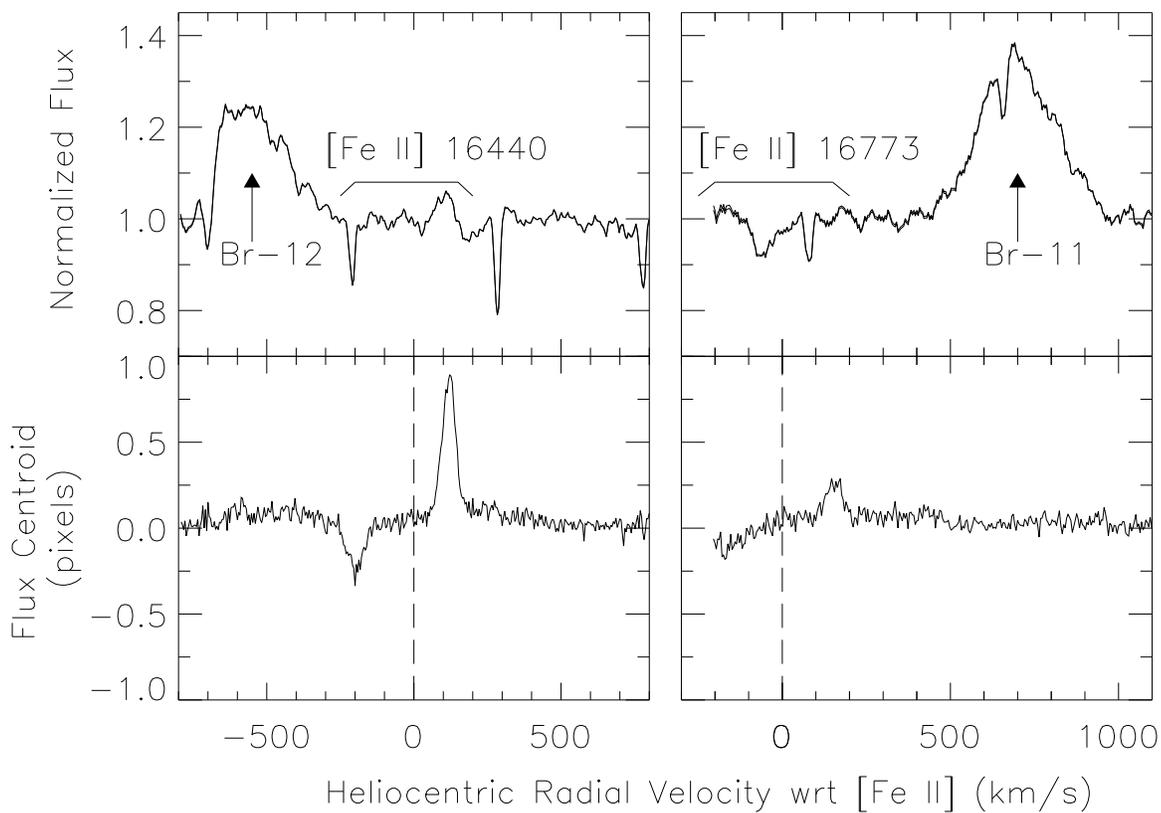}}
\caption{ Spectroastrometry of two [Fe~II] lines and two Brackett lines.
Top: stellar spectrum.  The [Fe II] lines at 1.644 $\mu$m and 1.677 $\mu$m
(extent of each line profile indicated by a horizontal
bracket) are overwhelmed by the stellar flux.
Bottom: spatial position of the flux centroid relative to that of the star.
Both Fe lines show clear signatures of bipolar outflow, while the Brackett lines 
are centered on the stellar position at all velocities.
The [Fe II] 1.677 $\mu$m line is located at the edge of an order and has an
uncertain velocity zero point.
}
\end{figure}

\vfill\eject
\begin{figure}
\vbox to 5in{\includegraphics{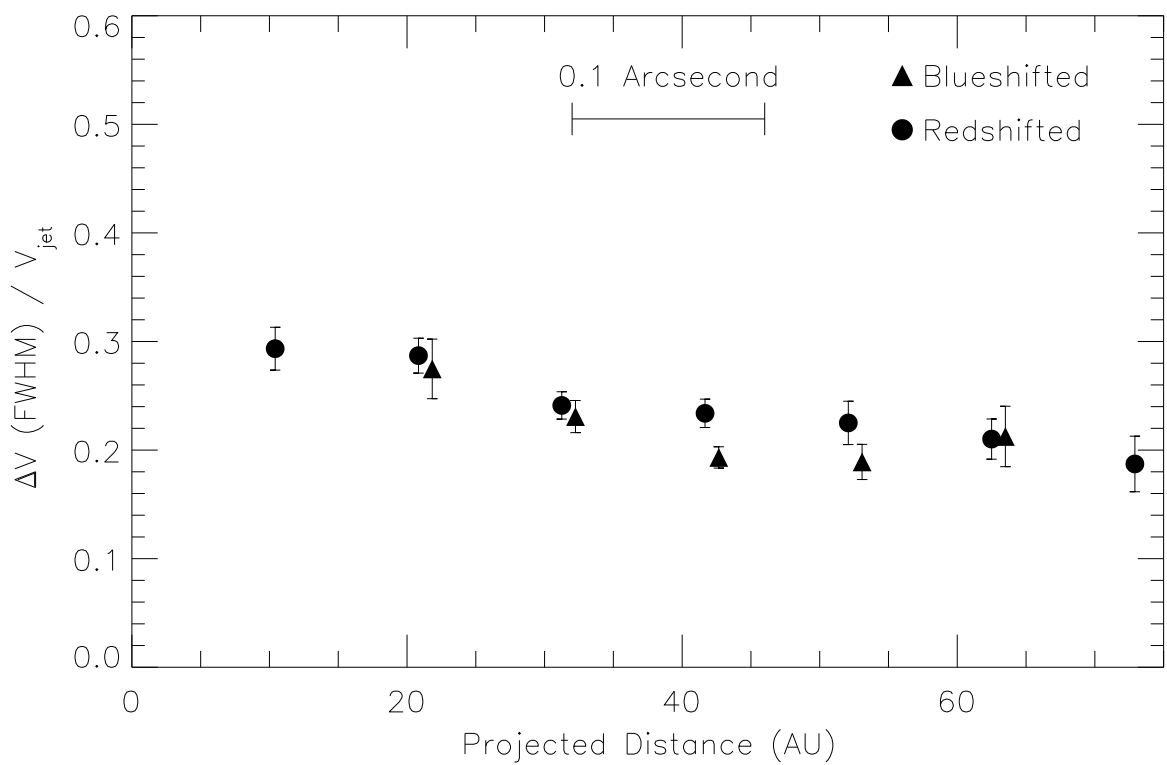}}
\caption{Ratio of the FWHM emission line width of [Fe~II] 1.644 $\mu$m, corrected
for instrumental broadening, to the jet velocity inferred from the radial velocities
in Fig.~2 and the known inclination angle of the flow. Points and errorbars for both the
redshifted and blueshifted jets derive from the on-axis values in Figs.~2 and 3. 
The expected ratio for radial flow with an opening angle of 4 degrees is $\sim$ 0.05,
so an extra source of line broadening is present in these flows. The redshifted and
blueshifted points are offset from one-another slightly in distance for clarity.
}
\end{figure}

\vfill\eject

\normalsize

\end{document}